\documentclass[aps,prd,amsmath,twocolumn,amssymbaps,showpacs]{revtex4-2}
\usepackage{graphicx}  
\usepackage{dcolumn}   
\usepackage{bm}        
\usepackage{amssymb}   
\usepackage{amsmath}   
\usepackage{bbm}
\usepackage{draftcopy}
\usepackage{relsize} 
\usepackage{xfrac}    
\usepackage{slashed}
\usepackage{color}
\usepackage{footnote}
\usepackage{changepage}
\usepackage{enumitem}
\definecolor{dgreen}{cmyk}{1.,0.,1.,0.2}        
\definecolor{orange}{cmyk}{0.,0.353,1.,0.}    

\usepackage[bookmarks]{hyperref}

\newcommand{\di}{{\rm d}}

\newcommand{\be}{\begin{equation}}
\newcommand{\ee}{\end{equation}}                                                                               
\newcommand{\bea}{\begin{eqnarray}}
\newcommand{\eea}{\end{eqnarray}} 

\begin{document}
\title{Charged pseudoscalar mesons in a strong magnetic field under the Weinberg model}
\author{Gaoqing Cao$^{1,2}$}
\affiliation{1 School of Physics and Astronomy, Sun Yat-sen University, Zhuhai 519082, China\\
2 Guangdong Provincial Key Laboratory of Quantum Metrology and Sensing, Sun Yat-Sen University, Zhuhai 519082 
China}
\date{\today}

\begin{abstract}
Recent lattice QCD simulations have further validated their earlier unusual findings: The lowest energies of charged pseudoscalar mesons $\pi^\pm$ and $K^\pm$ decrease at stronger magnetic field, though quasiparticle approximation assumes an increasing feature. We address this long-standing puzzle by employing the chiral effective Weinberg model, in which pseudoscalar and vector mesons exhibit intrinsic mutual couplings. Under this framework, charged pseudoscalar mesons deviate from pure quasiparticle behavior due to their interactions with neutral pseudoscalar and charged vector mesons. By incorporating the modifications induced by neutral pseudoscalar-charged vector loops, we demonstrate that the lowest energies of $\pi^\pm$ and $K^\pm$ indeed decrease at stronger magnetic field in both the lowest- and full-Landau-level calculations. However, instabilities emerge under a fixed mesonic coupling constant, and appear unavoidable when attempting to reproduce the observed peak structures. In contrast to the quark-antiquark meson description in models such as the NJL model, our results support the conjecture that a charged pseudoscalar meson could effectively form a molecular bound state of a neutral pseudoscalar meson and a charged vector meson in the strong magnetic field regime.
\end{abstract}


\begin{titlepage}
\maketitle
\end{titlepage}

\section{Introduction}
In high-energy nuclear physics, the modifications induced by extremely strong electromagnetic (EM) fields to the properties of quantum chromodynamics (QCD) matter represent a prominent topic of interest, driven by both experimental and theoretical motivations. Experimentally, EM fields with magnitudes comparable to the QCD energy scale are predicted to be generated in peripheral relativistic heavy-ion collisions~\cite{Skokov:2009qp,Deng:2012pc}, though their estimated lifetimes vary significantly across different theoretical frameworks~\cite{Shen:2025unr}. Topics such as the EM effect on the signals of chiral magnetic effect~\cite{Liao:2014ava,Kharzeev:2015kna,Huang:2015oca,Zhao:2022dac}, global polarizations of $\Lambda$ and $\bar{\Lambda}$~\cite{Li:2017slc,Guo:2019mgh,STAR:2021beb,STAR:2023nvo}, and electric charge correlations~\cite{Huang:2024hua,STAR:2023jdd} are widely explored. From the theoretical side, no sign problem arises in lattice QCD simulations for systems subject to a pure magnetic field, so abundant first-principle data can be provided to advance our understanding of the intrinsic properties of QCD matter. The most representative example is the observed "inverse magnetic catalysis effect"~\cite{Bali:2011qj,Bali:2012zg}: In contrast to the monotonically increasing behavior at low temperatures, the chiral condensate exhibits a decreasing trend with rising magnetic field once the temperature becomes sufficiently high. A straightforward and physically consistent explanation is that the effective coupling constant decreases with the magnetic field due to the intrinsic asymptotic freedom property of QCD~\cite{Ferrer:2014qka,Cao:2021rwx}.

Beyond the chiral condensate, the spectra of mesons also constitute a topic of substantial interest in a strong magnetic field: They not only encode key features of chiral symmetry~\cite{Klevansky:1992qe,Gusynin:1994xp,Miransky:2015ava}, but can also directly signal the emergence of instabilities to non-trivial phases of QCD matter~\cite{Chernodub:2010qx,Hidaka:2012mz,Bali:2017ian,Cao:2019res,Liu:2026kvs,Liu:2017spl,Cao:2019ctl,Chen:2019tcp,Cao:2020pmm,Cao:2015xja,Ke:2026npb}. Recently, building upon earlier efforts~\cite{Ding:2020hxw}, the in-magnetic-field spectra of both neutral and charged pseudoscalar mesons have been systematically investigated via lattice QCD simulations at an almost physical pion mass~\cite{Ding:2026qzu}. The state-of-the-art lattice QCD simulations further validated their earlier unusual findings: The lowest energies of charged pseudoscalar mesons $\pi^\pm$ and $K^\pm$ decrease at stronger magnetic field, though quasiparticle approximation assumes an increasing feature. The puzzle has remained unsolved for quite a long time until two related dedicated studies emerged this year: One attributes the observed decreasing behavior to strong magnetic field-induced $\pi^\pm-\rho^\pm$ mixing~\cite{Wang:2026xsm}, while the other interprets these features as the emergence of neutral pseudoscalar-charged vector molecular structure within the charged pseudoscalar mesons~\cite{Kojo:2026soi}. In our view, the latter interpretation is more natural, as the lowest energies of both neutral pseudoscalar and charged vector mesons are observed to decrease with the magnetic field~\cite{Bali:2017ian,Ding:2026qzu}. Furthermore, this interpretation is also more relevant, given that only the correlation functions between conjugate pseudoscalar fields were evaluated in the lattice QCD simulations~\cite{Ding:2026qzu}.

In Ref.~\cite{Kojo:2026soi}, the properties of charged pseudoscalar mesons are investigated within a nonrelativistic quantum mechanics framework based on the constituent quark picture, where issues associated with underlying internal symmetries are not explicitly incorporated. In this work, we address this long-standing puzzle by employing a quantum field theoretical approach -- the Weinberg model -- in which pseudoscalar and vector mesons exhibit intrinsic mutual couplings~\cite{Weinberg:1968de}. The Weinberg model was originally constructed to realize chiral symmetry nonlinearly~\cite{Weinberg:1968de}, making it well-suited for exploring the properties of light mesons that are sensitive to chiral symmetry dynamics~\cite{Klevansky:1992qe}. The work is arranged as follows: In Sec.~\ref{Weinberg model}, the frameworks of two- and three-flavor Weinberg models are established for a constant magnetic field in Sec.~\ref{two-flavor} and Sec.~\ref{three-flavor}, respectively. In Sec.~\ref{LLL}, we calculate the self energies by simply applying the lowest-Landau-level approximation, and then give a mathematical analysis on the origin of the observed decreasing behavior. The corresponding numerical results are presented in Sec.~\ref{LLLN} for illustration. In Sec.~\ref{FLL}, we calculate the self energies by adopting full Landau levels with the corresponding numerical results presented in Sec.~\ref{FLLN}. Finally, a summary is given in Sec.~\ref{summary}, followed by a detailed discussions on remaining issues.

\section{Weinberg model in a magnetic field}\label{Weinberg model}
\subsection{The two-flavor case}\label{two-flavor}
The chiral effective Weinberg model~\cite{Weinberg:1968de} is initially constructed to nonlinearly realize two-flavor chiral symmetry, the spirit of which eventually leads to establish chiral perturbation theory~\cite{Weinberg:1978kz,Gasser:1983yg} that is equivalent to quantum chromodynamics at low energy regime. As we know, the Weinberg model is itself a mesonic model with the lightest pseudoscalar pions and vector rhos the fundamental degrees of freedom thus is very suitable to explore the circumstances where these mesons are important. In the presence of a constant magnetic field, the original Lagrangian density can be extended by following the minimal coupling principle to be~\cite{Weinberg:1968de,Lee:1962vm}
\begin{eqnarray}
{\cal L}&=&{1\over 2}\left({(D_\mu\boldsymbol{\pi})^\dagger\cdot D^\mu\boldsymbol{\pi}\over(1+{\boldsymbol{\pi}^\dagger\cdot\boldsymbol{\pi}\over f_\pi^2})^2}-{m_\pi^2\,\boldsymbol{\pi}^\dagger\cdot\boldsymbol{\pi}\over 1+{\boldsymbol{\pi}^\dagger\cdot\boldsymbol{\pi}\over f_\pi^2}}\right)-{1\over 4} \boldsymbol{\rho}_{\mu\nu}^\dagger\cdot\boldsymbol{\rho}^{\mu\nu}\nonumber\\
&&+{m_\rho^2\over2}\left[\boldsymbol{\rho}_\mu\!+\!{g_\rho\boldsymbol{\pi}\!\times\! D_\mu\boldsymbol{\pi}\over m_\rho^2(1\!+\!{\boldsymbol{\pi}^\dagger\cdot\boldsymbol{\pi}\over f_\pi^2})}\right]^\dagger\cdot\left[\boldsymbol{\rho}^\mu\!+\!{g_\rho\boldsymbol{\pi}\!\times\! D^\mu\boldsymbol{\pi}\over m_\rho^2(1\!+\!{\boldsymbol{\pi}^\dagger\cdot\boldsymbol{\pi}\over f_\pi^2})}\right]\nonumber\\
&&\pm i{e\over 2}F^{\mu\nu}{\rho}_\mu^\mp{\rho}_\nu^\pm-{F^{\mu\nu}F_{\mu\nu}\over 4},
\end{eqnarray}
where the "factors" $\left(1+{\boldsymbol{\pi}^\dagger\cdot\boldsymbol{\pi}\over f_\pi^2}\right)^n\ ({n=-1,-2})$ are introduced to realize chiral symmetry, and the magnetic effect is encoded in the covariant derivative $D_\mu\equiv\partial_\mu+i\,qA_\mu$, rho meson strength tensors $\boldsymbol{\rho}_{\mu\nu}$, and electromagnetic strength tensor $F^{\mu\nu}$. 

Up to leading order, the "factors" can be taken to $1$, then the Lagrangian density simply becomes
\begin{eqnarray}\label{WM}
{\cal L}\!&=&\!-{F^{\mu\nu}F_{\mu\nu}\over 4}\!+\!{1\over 2}\left[{(D_\mu\boldsymbol{\pi})^\dagger\cdot D^\mu\boldsymbol{\pi}}\!-\!{m_\pi^2\,\boldsymbol{\pi}^\dagger\cdot\boldsymbol{\pi}}\right]\!-\!{\boldsymbol{\rho}_{\mu\nu}^\dagger\cdot\boldsymbol{\rho}^{\mu\nu}\over 4} \nonumber\\
&&+{m_\rho^2\over2}\boldsymbol{\rho}_\mu^\dagger\cdot\boldsymbol{\rho}^\mu\pm i{e\over 2}F^{\mu\nu}{\rho}_\mu^\mp{\rho}_\nu^\pm+{g_\rho}\boldsymbol{\rho}_\mu \cdot({\boldsymbol{\pi}\times i\,D^\mu\boldsymbol{\pi}}).
\end{eqnarray}
Here for convenience, the isovectors are defined in the electric charge eigenstates, that is, $\boldsymbol{\pi}=(\pi^0,\pi^-,\pi^+)$ and $\boldsymbol{\rho}=(\rho^0,\rho^-,\rho^+)$, and the latter is further defined in the spin eigenstates with $\boldsymbol{\rho}_\mu=(\boldsymbol{\rho}_t,\boldsymbol{\rho}_\downarrow,\boldsymbol{\rho}_0,\boldsymbol{\rho}_\uparrow)$~\cite{Cao:2020pmm}. Take $\rho$ mesons for example, the charge eigenstates are related to the isospin ones, $\rho^{i}\ (i=1,2,3)$, as
$$\rho^0=\rho^3,\ \rho^\pm={\rho^1\mp i\rho^2\over{\sqrt{2}}},$$
and the spin eigenstates are defined by the Lorentz components $\rho_a\ (a=t,x,y,z)$ as
$$\rho_0=\rho_z,\ \rho_{\uparrow/\downarrow}={\rho_{x}\mp i\,\rho_{y}\over \sqrt 2}.$$ 

The couplings between the electromagnetic field and rho mesons render the model non-renormalizable~\cite{Lee:1962vm}, and this property is somewhat subtle. In the vacuum, the strength tensors $\boldsymbol{\rho}_{\mu\nu}$ are defined in the same way as those for gauge fields in the $SU(2)$ Yang-Mills theory, that is,
$${\rho}_{\mu\nu}^a\equiv \partial_\mu{\rho}_{\nu}^a-\partial_\nu{\rho}_{\mu}^a+g_\rho\epsilon^{abc}{\rho}_{\mu}^b{\rho}_{\nu}^c\ \ (a,b,c=1,2,3)$$
with $\epsilon^{abc}$ the Levi-Civita symbol and the coupling constant $g_\rho=\sqrt{2}m_\rho/f_\pi$~\cite{Weinberg:1968de}. In a electromagnetic field, they should be defined through the charge eigenstates as~\cite{Cao:2020pmm}
\begin{eqnarray}
{\rho}_{\mu\nu}^0&=&\partial_\mu{\rho}^0_\nu-\partial_\nu{\rho}^0_\mu+i\,g_\rho (\rho^-_\mu\rho^+_\nu-\rho^-_\nu\rho^+_\mu),\\
{\rho}_{\mu\nu}^\pm&=&D_\mu^\pm{\rho}^\pm_\nu-D_\nu^\pm{\rho}^\pm_\mu,\ D_\mu^\pm\equiv\partial_\mu\pm ieA_\mu\mp i\,g_\rho{\rho}^0_\mu,
\end{eqnarray}
where ${\rho}_{\mu\nu}^\pm$ are effectively Abelian with a modified gauge field $A_\mu-{g_\rho\over e}{\rho}^0_\mu$.
In accordance with the spin eigenstates of rho mesons, the covariant derivatives should also be redefined as
\bea
D_0^{(\pm)}=D_z^{(\pm)},\ D_{\uparrow/\downarrow}^{(\pm)}=i{D_x^{(\pm)}\mp i\,D_y^{(\pm)}\over \sqrt 2}\label{SCD}
\eea
for all electric charge states. For the present study, we consider a constant magnetic field along $z$ direction, then the magnetic-spin coupling terms become explicitely~\cite{Lee:1962vm}
$$\pm i{e\over 2}F^{\mu\nu}{\rho}_\mu^\mp{\rho}_\nu^\pm=\pm {1\over 2}eB\left({\rho}_{\downarrow}^\mp{\rho}_\uparrow^\pm-{\rho}_{\uparrow}^\mp{\rho}_\downarrow^\pm\right).$$

In the following, we will choose the Landau gauge, $A_\mu=(0,By,0,0)$, for the constant magnetic field, then the momentum along $x$ direction is  well defined. To carry out calculations, it is more convenient to work in energy momentum and Landau spaces, so we will try to shift the Lagrangian to these spaces in the following. For that sake, neutral mesons can be expanded on the Fourier basis as
\bea
{\pi}^0(x)&=&\int {d^4k\over(2\pi)^4} e^{-i k\cdot x}{\pi}^0(k), \\
{\rho}_\mu^0(x)&=&\int {d^4k\over(2\pi)^4} e^{-i k\cdot x}{\rho}_\mu^0(k)
\eea
with ${k}\equiv (k_4,k_1,k_2,k_3)$, and charged mesons on the Ritus basis as~\cite{Miransky:2015ava,Ritus:1978cj}
\bea
{\pi}^\pm(x)&=&\sum_{n=0}\int {d^3 \tilde{k}\over(2\pi)^3} e^{\mp i \tilde{k}\cdot x}E_{\rm n}\left(y+{k_1\over |eB|}\right){\pi}^\pm_{n}({k}_\parallel), \\
{\rho}_\mu^\pm(x)&=&\sum_{n=0}\int {d^3 \tilde{k}\over(2\pi)^3} e^{\mp i \tilde{k}\cdot x}E_{\rm n}\left(y+{k_1\over |eB|}\right){\rho}_{\mu, n}^\pm({k}_\parallel)
\eea
 with $\tilde{k}\equiv (k_4,k_1,0,k_3), {k}_\parallel\equiv (k_4,0,0,k_3),$ and $E_{\rm n}(y)$ the $n$-th oscillation eigenfunction. Then, the diagonal kinetic parts of the Lagrangian can be found explicitly as~\cite{Cao:2020pmm}
\begin{widetext}
\begin{eqnarray}
{\cal L}_k&=&-{1\over 2}\int {d^4k\over(2\pi)^4}{{\pi}^0(-{k})\left(-k^2+m_{\pi^0}^2\right){\pi}^0({k})}-{1\over 2}\sum_{n=0}^\infty\int {d^3\tilde{k}\over(2\pi)^3}{{\pi}^\pm_{n}(k_\parallel)\left[k_4^2+(2n+1)|eB|+k_3^2+m_{\pi^\pm}^2\right]{\pi}^\mp_{n}(k_\parallel)}\nonumber\\
&&+{1\over 2} \int {d^4k\over(2\pi)^4}{\rho}^{0\mu}(-k)\left(-k^2+m_{\rho^0}^2\right){\rho}^{0}_{\mu}(k)+{1\over 2}\sum_{n=0}^\infty\int {d^3\tilde{k}\over(2\pi)^3}{\rho}^{\pm}_{t,n}(k_\parallel)\left[k_4^2+(2n+1)|eB|+{k}_3^2+m_{\rho^\pm_t}^2\right]{\rho}^{\mp}_{t,n}(k_\parallel)\nonumber\\
&&-{1\over 2}\sum_{n=0}^\infty\sum_{s=-1}^1\int {d^3\tilde{k}\over(2\pi)^3}{\rho}^{\pm}_{-s,n}(k_\parallel)\left[k_4^2+(2n+1\pm 2s)|eB|+{k}_3^2+m_{\rho^\pm_s}^2\right]{\rho}^{\mp}_{s,n}(k_\parallel)\label{Lk2}
\end{eqnarray}
with $k^2=-k_4^2-{\bf k}^2$.

For the interaction terms ${g_\rho}\boldsymbol{\rho}_\mu \cdot({\boldsymbol{\pi}\times i\,D^\mu\boldsymbol{\pi}})$, due to Lorentz contraction, the covariant derivatives must be shifted to the modified form \eqref{SCD} if the vector fields are defined as spin eigenstates. Consequently, the relevant derivatives can be evaluated as
\bea
i\,D^\mu{\pi}^0(x)&=&\int {d^4k\over(2\pi)^4} \bar{k}^\mu e^{-i k\cdot x}{\pi}^0({k}),\ \ \bar{k}^\mu\equiv-\left(k_4,{k_1-i\,k_2\over\sqrt{2}},k_3,{k_1+i\,k_2\over\sqrt{2}}\right);\\
i\,D^\mu_\parallel {\pi}^\pm(x)&=&\pm\sum_{n=0}\int {d^3 \tilde{k}\over(2\pi)^3} {k}_\parallel^\mu e^{\mp i \tilde{k}\cdot x}E_{\rm n}\left(y+{k_1\over |eB|}\right){\pi}^\pm_{n}({k}_\parallel),\\
i\,(D_x\mp i\,D_y) {\pi}^{(\pm)}(x)&=&(\mp)\sum_{n=0}\int {d^3 \tilde{k}\over(2\pi)^3} \sqrt{\left(n+{1\over2}\mp (\mp){1\over2}\right)|eB|} e^{(\mp) i \tilde{k}\cdot x}E_{\rm n\mp(\mp)1}\left(y+{k_1\over |eB|}\right){\pi}^{(\pm)}_{n}({k}_\parallel)
\eea
by utilizing the raising and lowering operator features of $D_x\mp i\,D_y$. Take $\rho^0_\mu$ and $\rho^-_\mu$ for example, the relevant interaction terms become
\bea
{\cal L}_{\rho^0\pi^-\pi^+}&\equiv&g_\rho \int d^4x \rho^0_\mu(x)\pi^-(x)i\,D^\mu \pi^+(x)={g_\rho}\sum_{n,n'=0}^\infty\int {d^4 {k}\over(2\pi)^4}\int {d^3 \tilde{p}\over(2\pi)^3} \Bigg[{p}_\parallel^\mu C_{n,n'}^{k_1,k_2,p_1} {\rho}_\mu^0({k}){\pi}^+_{n}({p}_\parallel){\pi}_{n'}^-({p}_\parallel^\mu+{k}_\parallel^\mu)\nonumber\\
&&\ \ \ \ \ \ \ \ \ \ \ \ \ \ \ \ +\sqrt{\left(n+{1\over2}\pm {1\over2}\right)|eB|}C_{n\pm1,n'}^{k_1,k_2,p_1}{\rho}_{\downarrow/\uparrow}^0({k}){\pi}^+_{n}({p}_\parallel){\pi}_{n'}^-({p}_\parallel^\mu+{k}_\parallel^\mu)\Bigg],\\
{\cal L}_{\rho^0\pi^+\pi^-}&\equiv&-g_\rho \int d^4x \rho^0_\mu(x)\pi^+(x)i\,D^\mu \pi^-(x)=-{g_\rho}\sum_{n,n'=0}^\infty\int {d^4 {k}\over(2\pi)^4}\int {d^3 \tilde{p}\over(2\pi)^3} \Bigg\{{p}_\parallel^\mu C_{n,n'}^{k_1,-k_2,p_1} {\rho}_\mu^0(-{k}){\pi}^-_{n}({p}_\parallel){\pi}_{n'}^+({p}_\parallel^\mu+{k}_\parallel^\mu)\nonumber\\
&&\ \ \ \ \ \ \ \ \ \ \ \ \ \ \ \ +\left[\sqrt{\left(n+{1\over2}\mp {1\over2}\right)|eB|}C_{n\mp1,n'}^{k_1,-k_2,p_1}\right]{\rho}_{\downarrow/\uparrow}^0(-{k}){\pi}^-_{n}({p}_\parallel){\rho}_{n'}^+({p}_\parallel^\mu+{k}_\parallel^\mu)\Bigg\},\\
{\cal L}_{\rho^-\pi^+\pi^0}&\equiv&g_\rho \int d^4x \rho^-_\mu(x)\pi^+(x)i\,D^\mu \pi^0(x)={g_\rho}\sum_{n,n'=0}^\infty\int {d^4 {k}\over(2\pi)^4}\int {d^3 \tilde{p}\over(2\pi)^3}\bar{k}^\mu C_{n,n'}^{k_1,k_2,p_1} {\pi}^0({k}){\pi}^+_{n}({p}_\parallel){\rho}_{\mu, n'}^-({p}_\parallel^\mu+{k}_\parallel^\mu),\\
{\cal L}_{\rho^-\pi^0\pi^+}&\equiv&-g_\rho \int d^4x \rho^-_\mu(x)\pi^0(x)i\,D^\mu \pi^+(x)=-{g_\rho}\sum_{n,n'=0}^\infty\int {d^4 {k}\over(2\pi)^4}\int {d^3 \tilde{p}\over(2\pi)^3} \Bigg[{p}_\parallel^\mu C_{n,n'}^{k_1,k_2,p_1} {\pi}^0({k}){\pi}^+_{n}({p}_\parallel){\rho}_{\mu, n'}^-({p}_\parallel^\mu+{k}_\parallel^\mu)\nonumber\\
&&\ \ \ \ \ \ \ \ \ \ \ \ \ \ \ \ +\sqrt{\left(n+{1\over2}\pm {1\over2}\right)|eB|}C_{n\pm1,n'}^{k_1,k_2,p_1}{\pi}^0({k}){\pi}^+_{n}({p}_\parallel){\rho}_{\downarrow/\uparrow, n'}^-({p}_\parallel^\mu+{k}_\parallel^\mu)\Bigg]
\eea
with the auxiliary function
\bea
C_{n,n'}^{k_1,k_2,p_1}&\equiv&\int \di y\ E_{\rm n}\left(y+{p_1\over |eB|}\right)E_{\rm n'}\left(y+{p_1+k_1\over |eB|}\right)e^{i\, k_2y}.
\eea
Then, the total interaction terms for $\rho^-$ are given by
\bea
{\cal L}_{\rho^-(\pi^+\pi^0)}
&\equiv&{\cal L}_{\rho^-\pi^+\pi^0}+{\cal L}_{\rho^-\pi^0\pi^+}=-{g_\rho}\sum_{n,n'=0}^\infty\int {d^4 {k}\over(2\pi)^4}\int {d^3 \tilde{p}\over(2\pi)^3} \Bigg\{({p}_\parallel^\mu-{k}_\parallel^\mu) C_{n,n'}^{k_1,k_2,p_1} {\pi}^0({k}){\pi}^+_{n}({p}_\parallel){\rho}_{\mu, n'}^-({p}_\parallel^\mu+{k}_\parallel^\mu)\nonumber\\
&&+\left[\sqrt{\left(n+{1\over2}\pm {1\over2}\right)|eB|}C_{n\pm1,n'}^{k_1,k_2,p_1}-{k_1\mp i\,k_2\over\sqrt{2}} C_{n,n'}^{k_1,k_2,p_1}\right]{\pi}^0({k}){\pi}^+_{n}({p}_\parallel){\rho}_{\downarrow/\uparrow, n'}^-({p}_\parallel^\mu+{k}_\parallel^\mu)\Bigg\},\label{rhom}
\eea
and the charge conjugate gives the one for $\rho^+$, that is,
\bea
{\cal L}_{\rho^+(\pi^0\pi^-)}
&\equiv&{\cal L}_{\rho^+\pi^0\pi^-}+{\cal L}_{\rho^+\pi^-0\pi^0}=-{g_\rho}\sum_{n,n'=0}^\infty\int {d^4 {k}\over(2\pi)^4}\int {d^3 \tilde{p}\over(2\pi)^3} \Bigg\{({p}_\parallel^\mu-{k}_\parallel^\mu) C_{n,n'}^{k_1,-k_2,p_1} {\pi}^0(-{k}){\pi}^-_{n}({p}_\parallel){\rho}_{\mu, n'}^+({p}_\parallel^\mu+{k}_\parallel^\mu)\nonumber\\
&&+\left[\sqrt{\left(n+{1\over2}\mp {1\over2}\right)|eB|}C_{n\mp1,n'}^{k_1,-k_2,p_1}-{k_1\pm i\,k_2\over\sqrt{2}}C_{n,n'}^{k_1,-k_2,p_1}\right]{\pi}^0(-{k}){\pi}^-_{n}({p}_\parallel){\rho}_{\downarrow/\uparrow, n'}^+({p}_\parallel^\mu+{k}_\parallel^\mu)\Bigg\}.\label{rhop}
\eea
In total, the Lagrangian follows as ${\cal L}={\cal L}_k+{\cal L}_{\rho^0\pi^\pm}+{\cal L}_{\rho^\pm(\pi^\mp\pi^0)}-{B^2\over 2}V_{4}$ with $V_4$ the space-time volume, ${\cal L}_{\rho^0\pi^\pm}\equiv{\cal L}_{\rho^0\pi^-\pi^+}+{\cal L}_{\rho^0\pi^+\pi^-},$ and ${\cal L}_{\rho^\pm(\pi^\mp\pi^0)}\equiv{\cal L}_{\rho^-(\pi^+\pi^0)}+{\cal L}_{\rho^+(\pi^0\pi^-)}$.

\subsection{Extension to the three-flavor case}\label{three-flavor}
To study the properties of $K^\pm$, pseudoscalars with strange/antistrange quarks, it is straightforward to extend the two-flavor case to the three-flavor case with nonet pseudoscalars $(\boldsymbol{\pi}, K^\pm, K^0, \bar{K}^0, \eta, \eta')$ and vectors $(\boldsymbol{\rho}, K^{*\pm}, K^{*0},\bar{K}^{*0}, \omega, \phi)$. The kinetic term can be directly modified from \eqref{Lk2} as
\begin{eqnarray}
{\cal L}_k^{\rm 3f}&=&\!-\!\!\sum_{S^0=\pi^0, K^0}^{\bar{K}^0, \eta, \eta'}\!{1\over 2}\int\! {d^4k\over(2\pi)^4}{S^0(-{k})\!\left(\!-k^2\!+\!m_{S^0}^2\right)\!S^0({k})}-\!\sum_{S^\pm=\pi^\pm}^{K^\pm}\!{1\over 2}\sum_{n=0}^\infty\int\! {d^3\tilde{k}\over(2\pi)^3}{{S}^\pm_{n}(k_\parallel)\!\left[k_4^2\!+\!(2n\!+\!1)|eB|\!+\!k_3^2\!+\!m_{S^\pm}^2\right]\!{S}^\mp_{n}(k_\parallel)}\nonumber\\
&&\!+\!\!\!\!\sum_{V^0=\rho^0,K^{*0}}^{\bar{K}^{*0}, \omega, \phi}\!\!{1\over 2}\! \int\!\! {d^4k\over(2\pi)^4}{V}^{0\mu}(-k)\!\left(\!-k^2\!+\!m_{V^0}^2\right)\!{V}^{0}_{\mu}(k)\!+\!\!\!\!\sum_{V^\pm=\rho^\pm}^{K^{*\pm}}\!\!\!{1\over 2}\sum_{n=0}^\infty\!\int\!\! {d^3\tilde{k}\over(2\pi)^3}{V}^{\pm}_{t,n}(k_\parallel)\!\left[k_4^2\!+\!(2n\!+\!1)|eB|\!+\!{k}_3^2\!+\!m_{V^\pm_t}^2\right]\!{V}^{\mp}_{t,n}(k_\parallel)\nonumber\\
&&-\sum_{V^\pm=\rho^\pm}^{K^{*\pm}}{1\over 2}\sum_{n=0}^\infty\sum_{s=-1}^1\int {d^3\tilde{k}\over(2\pi)^3}{V}^{\pm}_{-s,n}(k_\parallel)\left[k_4^2+(2n+1\pm 2s)|eB|+{k}_3^2+m_{V^\pm_s}^2\right]{V}^{\mp}_{s,n}(k_\parallel)\label{Lk3}.
\end{eqnarray}
For the two-flavor case, the interaction terms can be rewritten as 
\bea
{g_\rho}\boldsymbol{\rho}_\mu \cdot({\boldsymbol{\pi}\times i\,D^\mu\boldsymbol{\pi}})={g_\rho}\epsilon^{ijk}{\rho}_\mu^i {\pi}^ji\,D^\mu{\pi}^k=-{{g_\rho}\over \sqrt{2}}{\rm Tr}\ [({\rho}^i_\mu\tau^i{\pi}^j\tau^j-{\pi}^j\tau^j{\rho}^i_\mu\tau^i) i\,D^\mu{\pi}^k\tau^k]
\eea
 with $\boldsymbol{\tau}\equiv({\tau_3\over\sqrt{2}},{\tau_1-i\tau_2\over2},{\tau_1+i\tau_2\over2})$ defined in electric charge space. Then, the interaction terms in three-flavor case can be extended from that as $-{g_\rho\over\sqrt{2}}{\rm Tr}\ [({V}_\mu S-S {V}_\mu)i\,D^\mu S]$~\cite{Scherer:2002tk} with the pseudoscalar and vector matrices given by
\bea
S&\equiv& {1\over \sqrt{2}}\left(\eta{\lambda'^0}+\pi^0{\lambda^3}+\pi^\pm{\lambda^\pm_{12}}+K^\pm{\lambda^\pm_{45}}+K^0{\lambda^+_{67}}+\bar{K}^0{\lambda^-_{67}}+\eta'{\lambda'^8}\right),\nonumber\\
V&\equiv& {1\over \sqrt{2}}\left(\omega{\lambda'^0}+\rho^0{\lambda^3}+\rho^\pm{\lambda^\pm_{12}}+K^{*\pm}{\lambda^\pm_{45}}+K^{*0}{\lambda^+_{67}}+\bar{K}^{*0}{\lambda^-_{67}}+\phi{\lambda'^8}\right),
\eea
where $\lambda'^0\equiv{\rm diag}(1,1,0)$, $\lambda^\pm_{ij}\equiv {\lambda^i \pm i\,\lambda^j\over\sqrt{2}},$ and $\lambda'^8\equiv{\rm diag}(0,0,-\sqrt{2})$ with $\lambda^i (i=1,\dots,7)$ the Gell-Mann matrices. The interaction terms are effectively the same as that when we introduce the vector mesons as handed or handless $SU(3)$ gauge fields in the flavor space~\cite{Pisarski:1995xu}.

As the lowest energies of charged pseudoscalars increase with magnetic field when the latter is not large~\cite{Ding:2026qzu}, the contributions from exchanging neutral vector ($V^0$) currents is less important than the ones from exchanging charged vector ($V^\pm$) currents for the study of the charged pseudoscalars themselves. For simplicity, we suppress the interactions with $V^0$ here, then the interaction terms of $\boldsymbol{\pi}$ and the lowest-lying kaons, $K_0, \bar{K}_0$, and $K^\pm$, can be given explicitly as
\bea
&&{g_\rho}\left[{\rho}^-_\mu({\pi}^+ i\,D^\mu \pi^0-\pi^0 i\,D^\mu{\pi}^+)-{{K}^{*-}_\mu\over\sqrt{2}}({\pi}^+ i\,D^\mu K^0-K^0 i\,D^\mu{\pi}^+)+C.C\right]-{g_\rho\over \sqrt{2}}\bigg[{\rho}^-_\mu({K}^+ i\,D^\mu \bar{K}^0-\bar{K}^0 i\,D^\mu{K}^+)\nonumber\\
&&-{{K}^{*-}_\mu\over\sqrt{2}}(K^+ i\,D^\mu (\pi^0+\eta+\sqrt{2}\eta')-(\pi^0+\eta+\sqrt{2}\eta') i\,D^\mu K^+)+C.C.\bigg].
\eea
As we can see, if $n$ charged strange mesons are involved in a given term, the coupling constant would be suppressed by $2^{-n/2}$. Moreover, there are two kinds of interactions for $\pi^\pm$ but four kinds for $K^
\pm$, depending on whether the contributions of $\eta$ and $\eta'$ cancel out or not. Nevertheless, these extra contributions are relatively small compared to other terms, since the masses of $\eta, \eta',$ and ${K}^{*\pm}$ are relatively larger. If we check the interactions carefully, all the new terms actually can be obtained, up to the mentioned factors $2^{-n/2}$, from the two-flavor ones by just taking the substitutions $\rho^\mp\rightarrow {K}^{*\mp}$ and $\pi^0\rightarrow K^0$ for $\pi^\pm$, and $\rho^\mp\rightarrow {K}^{*\mp}$ or $\pi^0\rightarrow \bar{K}^0, \eta, \eta'$ for $\pi^\pm\rightarrow K^\pm$. In total, the relevant three-flavor Lagrangian to explore the properties of $\pi^\pm$ and $K^\pm$ is
\bea
{{\cal L}^{\rm 3f}}'={\cal L}_k^{\rm 3f}\!+\!{\cal L}_{\rho^\pm(\pi^\mp\pi^0)}\!-\!{1\over\sqrt{2}}{\cal L}_{{K}^{*\pm}(\pi^\mp K^0)}\!-\!{1\over\sqrt{2}}{\cal L}_{\rho^\pm(K^\mp \bar{K}^0)}\!+\!{1\over2}{\cal L}_{{K}^{*\pm}(K^\mp \pi^0))}\!+\!{1\over2}{\cal L}_{{K}^{*\pm}(K^\mp \eta)}\!+\!{1\over \sqrt{2}}{\cal L}_{{K}^{*\pm}(K^\mp \eta')}.
\eea

To study the properties of charged pseudoscalar mesons, the most important issue is to calculate their self-energies. In the three-flavor case, the self-energies of $\pi^\pm$ and $K^\pm$ can be calculated according to the Feynman diagrams given in Fig.~\ref{PKLE1}, the first of which is actually the one of $\pi^\pm$ in the two-flavor case.
\begin{figure}[!htb]
	\centering
	\includegraphics[width=0.85\textwidth]{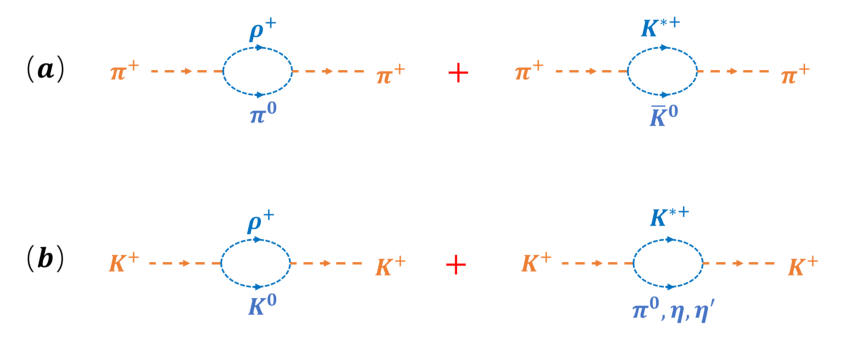}
	\caption{The Feynman diagrams for the self-energies of charged pion $\pi^+ (a)$ and kaon $K^+ (b)$ from the interactions with neutral pseudoscalars $S^0$ and charged vectors $V^+$.}\label{PKLE1}
\end{figure}
The following sections will be devoted to deriving these self-energies analytically by adopting the Lagrangian ${{\cal L}^{\rm 3f}}'$.

\section{The lowest-Landau-level approximation}\label{LLL}
To get a simple intuition, we first work with the lightest eigenstates of charged mesons in the lowest Landau level (LLL). To explore the LLL energy of $\pi^\pm_0$ for example, the lightest eigenstates of $\rho^\pm$ are ${\rho}_{\uparrow, 0}^+$ and ${\rho}_{\downarrow, 0}^-$ with the spins parallel and antiparallel to the magnetic field~\cite{Bali:2017ian}, respectively. According to \eqref{Lk2}, their eigenenergy is $E_{{\rho}_{\uparrow, 0}^+}=E_{{\rho}_{\downarrow, 0}^-}=\sqrt{{k}_3^2+M_{\rho^+_\uparrow0}^2}$ with the lowest energy square $M_{\rho^+_\uparrow0}^2\equiv -|eB|+m_{{\rho}_{\uparrow, 0}^+}^2$. In the LLL approximation, the relevant interaction terms in \eqref{rhom} and  \eqref{rhop} become simply
\bea
{\cal L}_{\rho^\pm(\pi^\mp\pi^0)}&=&-{g_\rho}\int {d^4 {k}\over(2\pi)^4}\int {d^3 \tilde{p}\over(2\pi)^3} \Bigg\{\left[\sqrt{|eB|}C_{1,0}^{k_1,k_2,p_1}-{k_1- i\,k_2\over\sqrt{2}} C_{0,0}^{k_1,k_2,p_1}\right]{\pi}^0({k}){\pi}^+_{0}(p_\parallel){\rho}_{\downarrow, 0}^-({p}_\parallel^\mu+{k}_\parallel^\mu)\nonumber\\
&&+\left[\sqrt{|eB|}C_{1,0}^{k_1,-k_2,p_1}-{k_1+ i\,k_2\over\sqrt{2}}C_{0,0}^{k_1,-k_2,p_1}\right]{\pi}^0(-{k}){\pi}^-_{0}(p_\parallel){\rho}_{\uparrow, 0}^+({p}_\parallel^\mu+{k}_\parallel^\mu)\Bigg\}.
\eea
The auxiliary functions $C_{0,0}^{k_1,k_2,p_1}$ and $C_{1,0}^{k_1,k_2,p_1}$ can be worked out explicitly as
\bea
C_{0,0}^{k_1,k_2,p_1}&\equiv& \left({|eB|\over\pi}\right)^{1/2}\int_{-\infty}^\infty \di y\, e^{-{|eB|\over2}\left[\left(y+{p_1\over |eB|}\right)^2+\left(y+{p_1+k_1\over |eB|}\right)^2\right]+i\,k_2y}=e^{-{1\over4|eB|}\left[k_2^2+k_1^2+2i\,k_2(2p_1+k_1)\right]},\\
C_{1,0}^{k_1,k_2,p_1}&\equiv& \left({2\over\pi}\right)^{1/2}|eB|\int_{-\infty}^\infty \di y\, \left(y+{p_1\over |eB|}\right)e^{-{|eB|\over2}\left[\left(y+{p_1\over |eB|}\right)^2+\left(y+{p_1+k_1\over |eB|}\right)^2\right]+i\,k_2y}=-{k_1-i\,k_2\over \sqrt{2|eB|}}C_{0,0}^{k_1,k_2,p_1}.\label{Cs}
\eea
Actually, the latter is consistent with the intuitive expectation when one takes a partial integration to $\int d^4x \rho^-_\mu(x)\pi^0(x)i\,D^\mu \pi^+(x)$ and applies the gauge condition $D^\mu\boldsymbol{\rho}_\mu=0$. Eventually, the interaction terms reduce to a very simple form,
\bea
{\cal L}_{\rho^\pm(\pi^\mp\pi^0)}&=&{2g_\rho}\int {d^4 {k}\over(2\pi)^4}\int {d^3 \tilde{p}\over(2\pi)^3} \left[{k_1- i\,k_2\over\sqrt{2}} C_{0,0}^{k_1,k_2,p_1}{\pi}^0({k}){\pi}^+_{0}(p_\parallel){\rho}_{\downarrow, 0}^-({p}_\parallel^\mu+{k}_\parallel^\mu)\right.\nonumber\\
&&\left.\ \ \ \ \ \ \ \ \ \ \ \ \ \ \ \ \ \ \ \ \ \ \ \ \ \ \ \ \ \ \ \ \ \ \ \ \ \ \ \ \ \ \ \ \  \ \ \ \ \ \ \ \ \ \ \ \ +{k_1+ i\,k_2\over\sqrt{2}}C_{0,0}^{k_1,-k_2,p_1}{\pi}^0(-{k}){\pi}^-_{0}(p_\parallel){\rho}_{\uparrow, 0}^+({p}_\parallel^\mu+{k}_\parallel^\mu)\right]. \label{int2}
\eea

Now, we are ready to calculate the first Feynman diagram in Fig.~\ref{PKLE1} (a) by putting in the interaction vertices in \eqref{int2} and the propagators of $\pi^0({k})$ and ${\rho}_{\downarrow, 0}^-({k}_\parallel)$ in \eqref{Lk3}, that is,
\bea
\Pi_{\pi^\pm}({p}_\parallel)&=&-\left({2g_\rho}\right)^2\int {\di^4 {k}\over(2\pi)^4}{k_1- i\,k_2\over\sqrt{2}} C_{0,0}^{k_1,k_2,p_1}{k_1+ i\,k_2\over\sqrt{2}}C_{0,0}^{k_1,-k_2,p_1}{1\over -k^2+m_{\pi^0}^2}{1\over -({p}_\parallel+{k}_\parallel)^2+M_{\rho^+_\uparrow0}^2}\nonumber\\
&=&-\left({2g_\rho}\right)^2\int {\di^4 {k}\over(2\pi)^4}{k_\bot^2\over2} e^{-{k_\bot^2\over2|eB|}}{1\over -k^2+m_{\pi^0}^2}{1\over -({p}_\parallel+{k}_\parallel)^2+M_{\rho^+_\uparrow0}^2}.\label{SEpi0}
\eea
By completing the integration over energy $k_4$ first, we have
\bea
\Pi_{\pi^\pm}({p}_\parallel)&=&-\sum_{t=\pm}\left({2g_\rho}\right)^2\int {\di^3 {k}\over(2\pi)^3}{k_\bot^2\over2} e^{-{k_\bot^2\over2|eB|}}\left[{1\over 4E_{\pi^0}}{1\over (p_4+i\,t\,E_{\pi^0})^2+E_{\rho^+_\uparrow0}^2}+{1\over 4E_{\rho^+_\uparrow0}}{1\over (p_4-i\,t\,E_{\rho^+_\uparrow0})^2+E_{\pi^0}^2}\right]\nonumber\\
&=&=-\left({2g_\rho}\right)^2\int {\di^3 {k}\over(2\pi)^3}{k_\bot^2\over2} e^{-{k_\bot^2\over2|eB|}}{E_{\pi^0}^{-1}+E_{\rho^+_\uparrow0}^{-1}\over 2[(E_{\pi^0}+E_{\rho^+_\uparrow0})^2+p_4^2]};\label{SEpi1}
\eea
by completing the integrations over transverse momenta $k_\bot$ first, we have 
\bea
\Pi_{\pi^\pm}({p}_\parallel)&=&-|eB|{g_\rho^2\over\pi}\int {\di^2 {k}_\parallel\over(2\pi)^2}{1+\tilde{E}_\parallel e^{\tilde{E}_\parallel} {\rm Ei} (-\tilde{E}_\parallel)\over -({p}_\parallel+{k}_\parallel)^2+M_{\rho^+_\uparrow0}^2}\label{SEpi2}
\eea
with $\tilde{E}_\parallel\equiv{-k_\parallel^2+m_{\pi^0}^2\over 2|eB|}$ and ${\rm Ei} (x)$ the exponential integral function. The former form is more convenient for numerical calculations, while the latter one is more useful for analytical explorations. As $\lim_{\tilde{E}_\parallel\rightarrow\infty}[1+\tilde{E}_\parallel e^{\tilde{E}_\parallel} Ei(-\tilde{E}_\parallel)]=0$, it is straightforward to verify that the integrations over the longitudinal momentum ${k}_\parallel$ are finite, so no regularization is needed in the LLL approximation, and the self-energy also vanishes in the zero magnetic field limit. These nontrivial findings are highly useful for obtaining regularization-independent results, even though the Weinberg model with fundamental charged vectors is non-renormalizable in an external magnetic field~\cite{Lee:1962vm}. According to numerical estimations, the effect of ${p}_\parallel$ is not important here $(< 10\%)$ for $|{p}_\parallel|\lesssim|eB|$, thus we can simply take $\Pi_{\pi^\pm}({p}_\parallel)\approx \Pi_{\pi^\pm}(0)=-|eB|\ \xi(|eB|)$ in this case, where the dimensionless auxiliary function is
\bea
\xi(|eB|)&\equiv& {g_\rho^2\over\pi}\int {\di^2 {k}_\parallel\over(2\pi)^2}{1+\tilde{E}_\parallel e^{\tilde{E}_\parallel} {\rm Ei} (-\tilde{E}_\parallel)\over -{k}_\parallel^2+M_{\rho^+_\uparrow0}^2}={g_\rho^2\over4\pi^2}\int_{m_{\pi^0}^2/|2eB|}^\infty {\di y}{1+y\, e^{y}\, {\rm Ei} (-y)\over y+({M}_{\rho^+_\uparrow0}^2-{m}_{\pi^0}^2)/|2eB|}.\label{xi}
\eea

Recalling the free propagator of $\pi^\pm_0({k}_\parallel)$ in \eqref{Lk3}, the inverse propagator is corrected by the self-energy to $D_{\pi^\pm_0}^{-1}=-p_\parallel^2+|eB|+m_{\pi^\pm}^2+\Pi_{\pi^\pm}({p}_\parallel)$, and the lowest energy $p_0$ is to be solved self-consistently from the equation
\bea
-p_0^2+|eB|+m_{\pi^\pm}^2+\Pi_{\pi^\pm}(i\,p_0)=0\label{SCEpi}
\eea
for $p_3=0$. Then, for $p_0\lesssim|eB|$, the lowest energy square of $\pi^\pm$ can be simply derived as $M^2_{\pi^\pm_0}=[1-\xi(|eB|)]|eB|+m_{\pi^\pm}^2$, and the coefficient $1-\xi(|eB|)$ determines if the energy increases or decreases with $|eB|$. If $|eB|$ is very large, ${M}_{\rho^+_\uparrow0}^2/|2eB|, {m}_{\pi^0}^2/|2eB|\rightarrow 0$, then the integration in \eqref{xi} is dominated by the region $y\sim 0$. As $\lim_{y\rightarrow 0}y\, e^{y}\, {\rm Ei} (-y)=0$, the integration can be estimated as $\xi(|eB|)\approx {g_\rho^2\over4\pi^2}\ln (|2eB|/{M}_{\rho^+_\uparrow0}^2)$ and $1-\xi(|eB|)<0$ for large $eB$. So, the lowest energy of $\pi^\pm$ is found to decrease with large $eB$, qualitatively consistent with the lattice results.
Nevertheless, another problem emerges here: the lowest energy square could be negative at very large $eB$, indicating an unexpected phase of pion superfluidity. In contrast, the results from quark models are rather different for $\pi^\pm$~\cite{Cao:2015xja,Ke:2026npb}: The leading contributions are from the loops composed of the LLL quark and first Landau level antiquark or verse vice, that is, 
\bea
\Pi_{\pi^\pm}({p}_\parallel=0)=8a\,N_c|eB|\sum_{n=0,1}\int {\di^2 {k}_\parallel\over(2\pi)^2}{(-{k}_\parallel^2+m_um_d)\over (-{k}_\parallel^2+2n|q_uB|+m_{u}^2) (-{k}_\parallel^2+2(1-n)|q_dB|+m_{d}^2)}
\eea
with the positive coefficient $a$ depending on the choice of the eigenstate of $\pi^\pm$. Though the term is divergent and needs further regularization, its key characteristics with respect to $eB$ remain unchanged. As a result, the lowest energy of $\pi^\pm$ is generally positive and increases monotonically with $eB$ in quark models.

In the three-flavor case, all the loops in Fig.~\ref{PKLE1} contribute and the self-energies of $\pi^\pm$ and $K^\pm$ follow, respectively, as
\bea
\Pi_{\pi^\pm}({p}_\parallel)
&=&-\left({2g_\rho}\right)^2\int {\di^3 {k}\over(2\pi)^3}{k_\bot^2\over2} e^{-{k_\bot^2\over2|eB|}}\left[{E_{\pi^0}^{-1}+E_{\rho^+_\uparrow0}^{-1}\over 2\left((E_{\pi^0}+E_{\rho^+_\uparrow0})^2+p_4^2\right)}+{E_{K^0}^{-1}+E_{K^{*+}_\uparrow0}^{-1}\over4\left( (E_{K^0}+E_{K^{*+}_\uparrow0})^2+p_4^2\right)}\right],\label{SEpi3f}\\
\Pi_{K^\pm}({p}_\parallel)
&=&-\left({2g_\rho}\right)^2\int {\di^3 {k}\over(2\pi)^3}{k_\bot^2\over2} e^{-{k_\bot^2\over2|eB|}}\left[{E_{K^0}^{-1}+E_{\rho^+_\uparrow0}^{-1}\over 4\left(E_{K^0}+E_{K^{*+}_\uparrow0})^2+p_4^2\right)}+\sum_{S^0=\pi^0}^{\eta,\eta'}{\alpha_{S^0}(E_{S^0}^{-1}+E_{K^{*+}_\uparrow0}^{-1})\over 8\left(E_{S^0}+E_{K^{*+}_\uparrow0})^2+p_4^2\right)}\right]\label{SEK3f}
\eea
with $\alpha_{\eta'}=2\alpha_{\pi^0}=2\alpha_\eta=2$, ${E}_{S^0}=\sqrt{{k}_3^2+y+m_{S^0}^2}\ (S^0=\pi^0, \eta,\eta', K^0),$ and ${E}_{V^+_\uparrow0}\equiv \sqrt{{k}_3^2+M_{V_\uparrow^+0}^2} \ (V^+_\uparrow=\rho^+_\uparrow, K^{*+}_\uparrow)$. Compared to \eqref{SEpi1}, all the new terms are the same except for the involved meson masses, so the self-energy of $\pi^\pm$ would decrease more quickly in the three-flavor case. Similar to the equation \eqref{SCEpi} for $\pi^\pm$, the lowest energy $p_0$ of $K^\pm$ is to be solved self-consistently from the equation
\bea
-p_0^2+|eB|+m_{K^\pm}^2+\Pi_{K^\pm}(i\,p_0)=0\label{SCEK}
\eea
for $p_3=0$. If the neutral pseudoscalars and charged vectors are separately degenerate in the magnetic field, we will found $\Pi_{\pi^\pm}({p}_\parallel)=\Pi_{K^\pm}({p}_\parallel)$ as should be, that is, the degeneracy of $\pi^\pm$ and $K^\pm$ in the vacuum will not be broken by the loops in the magnetic field.
\end{widetext}

\subsection{Numerical results} \label{LLLN}
\begin{figure}[!htb]
	\centering
	\includegraphics[width=0.42\textwidth]{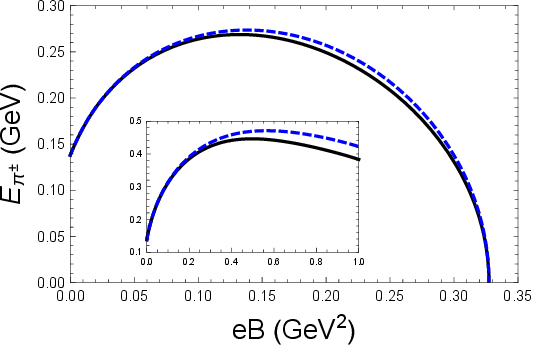}
	\caption{The lowest energy of $\pi^\pm$, $E^2_{\pi^\pm}$, as functions of $eB$ from Eq.~\eqref{SCEpi} (black solid) and the approximation $M_{\pi^\pm_0}$ (blue dashed). The insert corresponds to $g_\rho=4$.}\label{PLE}
\end{figure}

To carry out numerical calculations, the effective coupling constant was found to be $g_\rho=6.07$ in order to reproduce the $\rho^0$ decay width $\Gamma_\rho=155~{\rm GeV}$ in vacuum~\cite{Gale:1990pn,Cao:2025xud}. Usually, all the masses of the involved mesons are simply given by their vacuum values found in experiments, and the effective masses under a special condition can be studied by taking into account self-energy corrections from quantum loops. To calculate the self-energies of charged pseudoscalars in a magnetic field, we consider only one kind of loops that are composed of the propagators of neutral pseudoscalars and charged vectors. Then, to be more consistent with real QCD, we will adopt, in a quenched manner, the $|eB|$-dependent lowest energies of neutral pseudoscalars and charged vectors given by lattice QCD simulations~\cite{Bali:2017ian,Ding:2026qzu}, that is, $m_{\pi^0}(eB), m_{K^0}(eB), m_{\eta}(eB),$ and ${M}_{\rho^+_\uparrow0}(eB)$. There are no lattice data on the lowest energies of $\eta'$ and $ K^{*+}_\uparrow$ yet, so we will simply assume that their ratios to vacuum masses follow those of $\eta$ and $\rho^+_{\uparrow}$, respectively. Obviously, with these inputs from lattice QCD, the artificial vacuum superconductivity can be well avoided at tree level.

\begin{figure}[!htb]
	\centering
	\includegraphics[width=0.42\textwidth]{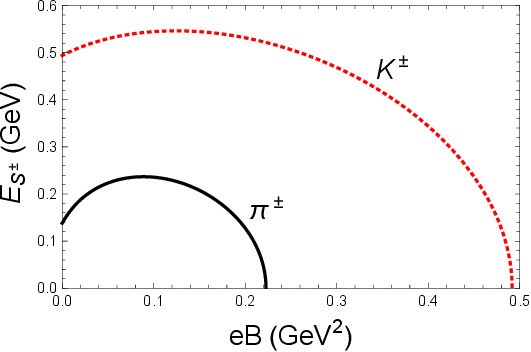}
	\caption{The lowest energies of $\pi^\pm$ and $K^\pm$, $E^2_{\pi^\pm}$ (black solid) and $E^2_{K^\pm}$ (red dotted), as functions of $eB$ with the lowest-Landau-level self-energies Eq.~\eqref{SEpi3f} and Eq.~\eqref{SEK3f}, respectively.}\label{PKLE}
\end{figure}

First of all, we compare the lowest energy squares of $\pi^\pm$ given self-consistently by Eq.~\eqref{SCEpi} and the approximate $M^2_{\pi^\pm_0}$ in Fig.~\ref{PLE}. Indeed, they match each other quite well hence justify the previous qualitative discussions with $M^2_{\pi^\pm_0}$, that is, the lowest energy of $\pi^\pm$ would decrease with larger eB. But the turn point is around $eB=0.14~{\rm GeV}^2$ compared to $0.6~{\rm GeV}^2$ from lattice QCD~\cite{Ding:2026qzu}, the main reason for the quantitative inconsistency is that the LLL approximation to $\rho^+_\uparrow$ is not valid at small $eB$. As mentioned before, $M^2_{\pi^\pm_0}$ would decrease across zero to be negative at larger $eB$, in contradiction to the lattice results~\cite{Ding:2026qzu}. As we usually expect the LLL approximation to be valid at very large $eB$, such a nonphysical problem cannot be solved by adopting full Landau level   propagator for $\rho^+_\uparrow$. Neither can the ignored loops with neutral vectors help since the contributions are of the same sign as those in Fig.~\ref{PKLE}. Nevertheless, the problem might be cured by simply taking into account a running coupling constant $g_\rho(eB)$ within Weinberg model: If we take $g_\rho=4$ for example, the change point would be shifted to around $eB=0.5~{\rm GeV}^2$ and the instability is greatly postponed, see the insert in Fig.~\ref{PLE}. 

In three-flavor case, the lowest energies of $\pi^\pm$ and $K^\pm$ are demonstrated in Fig.~\ref{PKLE} for the fixed coupling constant. As we can see, the lowest energy of $\pi^\pm$ keeps the same feature but with the peak shifting to smaller $eB$, and that of $K^\pm$ shares similar features.

\begin{widetext}
\section{Full-Landau-level calculations}\label{FLL}
As we know, the LLL approximation is usually not so good when the magnetic field is not so large. To explore the properties of $\pi^\pm$ and $K^\pm$ over the whole magnetic field range, we had better to adopt the full Landau level forms for the charged vectors. Still, take $\pi^\pm$ for example and first focus on the first Feynman diagram in Fig.~\ref{PKLE1} $(a)$. Inspired by the relation in \eqref{Cs}, the interaction terms in ${\cal L}_{\rho^\pm(\pi^\mp\pi^0)}$ can be rewritten in more compact forms as
\bea
{\cal L}_{\rho^\pm(\pi^\mp\pi^0)}=2{g_\rho} \int d^4x \rho^-_\mu(x)\pi^+(x)i\,D^\mu \pi^0(x)-2{g_\rho} \int d^4x \rho^+_\mu(x)\pi^-(x)i\,D^\mu \pi^0(x)
\eea
by taking partial integrals and applying Landau gauge to the charged vectors, that is, $D^\mu\rho^\pm_\mu(x)=0$. Then, to study the lowest energy of $\pi^\pm$, the interaction terms can be rewritten in energy momentum and Landau spaces as
\bea
{\cal L}_{\rho^\pm(\pi^\mp\pi^0)}&=&-2{g_\rho}\sum_{n=0}^\infty\int {d^4 {k}\over(2\pi)^4}\int {d^3 \tilde{p}\over(2\pi)^3} \Bigg\{{k_1- i\,k_2\over\sqrt{2}} C_{0,n}^{k_1,k_2,p_1}{\pi}^0({k}){\pi}^+_{0}({p}_\parallel){\rho}_{\downarrow, n}^-({p}_\parallel^\mu+{k}_\parallel^\mu)\nonumber\\
&&\ \ \ \ \ \ \ \ \ \ \ \ \ \ \ \ \ \ \ \ \ \ \ \ \ \ \ \ \ \ \ \ \ \ \ \ \ \ \ \ \ \ \ \ \  \ \ \ \ \ \ \ \ \ \ \ \ -{k_1+ i\,k_2\over\sqrt{2}}C_{0,n}^{k_1,-k_2,p_1}{\pi}^0(-{k}){\pi}^-_{0}({p}_\parallel){\rho}_{\uparrow, n}^+({p}_\parallel^\mu+{k}_\parallel^\mu)\Bigg\},
\eea
where all the Landau levels of ${\rho}_{\uparrow/\downarrow}^\pm$ have been taken into account. The explicit form of $C_{0,n}^{k_1,k_2,p_1}$ can be calculated as
\bea
C_{0,n}^{k_1,k_2,p_1}&\equiv& \left({|eB|\over\pi 2^n n!}\right)^{1/2}\int_{-\infty}^\infty \di y\, e^{-{|eB|\over2}\left[\left(y+{p_1\over |eB|}\right)^2+\left(y+{p_1+k_1\over |eB|}\right)^2\right]+i\,k_2y}H_{n}\left(\sqrt{|eB|}\left(y+{p_1+k_1\over |eB|}\right)\right)\nonumber\\
&=&\left({1\over n!}\right)^{1/2}e^{-{1\over4|eB|}\left[k_2^2+k_1^2+2i\,k_2(2p_1+k_1)\right]}\left({k_1+i\,k_2\over \sqrt{2|eB|}}\right)^n,
\eea
hence the self-energy of $\pi^\pm$ follows the first Feynman diagram in Fig.~\ref{PKLE1} $(a)$ as
\bea
\Pi_{\pi^\pm}({p}_\parallel)&=&-\left({2g_\rho}\right)^2\sum_{n=0}\int {\di^4 {k}\over(2\pi)^4}{k_1- i\,k_2\over\sqrt{2}} C_{0,n}^{k_1,k_2,p_1}{k_1+ i\,k_2\over\sqrt{2}}C_{0,n}^{k_1,-k_2,p_1}{1\over -k^2+m_{\pi^0}^2}{1\over -({p}_\parallel+{k}_\parallel)^2+(2n-1)|eB|+m_{\rho^+_\uparrow0}^2}\nonumber\\
&=&-{\left({2g_\rho}\right)^2}\sum_{n=0}{1\over n!}\int {\di^4 {k}\over(2\pi)^4}{k_\bot^2\over2} \left({k_\bot^2\over 2|eB|}\right)^n
e^{-{k_\bot^2\over2|eB|}}{1\over -k^2+m_{\pi^0}^2}{1\over -({p}_\parallel+{k}_\parallel)^2+(2n-1)|eB|+m_{\rho^+_\uparrow0}^2}\nonumber\\
&=&-{\left({2g_\rho}\right)^2}\sum_{n=0}{1\over n!}\int_0^\infty \di s\int {\di^4 {k}\over(2\pi)^4}{k_\bot^2\over2} 
\left({k_\bot^2\over 2|eB|}\right)^ne^{-{k_\bot^2\over2|eB|}}e^{-s\big(-({p}_\parallel+{k}_\parallel)^2+(2n-1)|eB|+m_{\rho^+_\uparrow0}^2\big)}{1\over -k^2+m_{\pi^0}^2}\nonumber\\
&=&-{\left({2g_\rho}\right)^2}\int_0^\infty \di s\int {\di^4 {k}\over(2\pi)^4}{k_\bot^2\over2} 
e^{-s\big(-({p}_\parallel+{k}_\parallel)^2-|eB|+{1-e^{-2|eB|s}\over 2|eB|s}k_\bot^2+m_{\rho^+_\uparrow0}^2\big)}{1\over -k^2+m_{\pi^0}^2}.\label{SEpiF1}
\eea
In the last two steps, we have introduced a proper-time integral to represent the summation over Landau levels $n$. In the limit $eB\rightarrow 0$, the integration over $s$ can be completed in the last step, and the result well reduces to the transverse part of the vacuum term without external magnetic field.

Alternatively, as the Schwinger phases cancel out in the calculations, one can use the vertices attached to $\pi^\pm_0$ and the propagator of ${\rho}_{\uparrow/\downarrow}^\pm$ in energy momentum space to get the same result.  Actually, the Schwinger phases are involved in the eigenstate correlation of $\pi^\pm_0$ and the propagator of ${\rho}_{\uparrow/\downarrow}^\pm$, that is,
\bea
V_{\pi^\pm_0}(x,x')&\equiv& \int {d^3 \tilde{k}\over(2\pi)^3}e^{- i \tilde{k}\cdot x}E_{\rm 0}\left(y+{k_1\over |eB|}\right)e^{ i \tilde{k}\cdot x'}E_{\rm 0}\left(y'+{k_1\over |eB|}\right), \\D_{\rho^+_\uparrow}(x,x')&\equiv&\sum_{n=0}^\infty\int {d^3 \tilde{k}\over(2\pi)^3}{e^{- i \tilde{k}\cdot x}E_{\rm n}\left(y+{k_1\over |eB|}\right)e^{ i \tilde{k}\cdot x'}E_{\rm n}\left(y'+{k_1\over |eB|}\right)\over -{k}_\parallel^2+(2n-1)|eB|+m_{\rho^+_\uparrow}^2}.
\eea
Then, after getting rid of the Schwinger phase, Fourier transformations can be performed to give~\cite{Miransky:2015ava,Kuznetsov:2015uca}
\bea
V_{\pi^\pm_0}(k_\bot)=2e^{-k^2_\bot/|eB|},\ D_{\rho^+_\uparrow}(k)=\int \di s\, {\rm sech} (|eB|s) e^{-s\big[-{k}_\parallel^2-2|eB|+{\tanh(|eB|s)\over |eB|s}{k}_\bot^2+m_{\rho^+_\uparrow}^2\big]}
\eea
with the latter in Schwinger representation. Then, the integration over the transverse momenta of $\pi^\pm$ gives
\bea
&&\int{\di^2p_\bot\over(2\pi)^2}V_{\pi^\pm_0}(p_\bot)D_{\rho^+_\uparrow}(p+k)=\int {\di s\over 2\pi} {|eB|{\rm sech} (|eB|s)\over 1+\tanh(|eB|s)} e^{-s\big[-({p}_\parallel+{k}_\parallel)^2-2|eB|+{\tanh(|eB|s)\over |eB|s(1+\tanh(|eB|s))}k_\bot^2+m_{\rho^+_\uparrow}^2\big]}\nonumber\\
&=&|eB|\int {\di s\over 2\pi} e^{-s\big[-({p}_\parallel+{k}_\parallel)^2-|eB|+{\tanh(|eB|s)\over |eB|s(1+\tanh(|eB|s))}k_\bot^2+m_{\rho^+_\uparrow}^2\big]}=|eB|\int {\di s\over 2\pi} e^{-s\big[-({p}_\parallel+{k}_\parallel)^2-|eB|+{1-e^{-2|eB|s}\over 2|eB|s}k_\bot^2+m_{\rho^+_\uparrow}^2\big]},\label{VD}
\eea
and the self-energy is just ${|eB|\over 2\pi}$ times that given in \eqref{SEpiF1}. Recall that the integration over $p_1$ of $\pi^\pm$ in \eqref{Lk3} would also give a Landau degeneracy factor ${|eB|\over 2\pi}$, the modification to the dispersion is essentially the same as that of \eqref{SEpiF1} regardless $p_1$ is integrated over or not.

We now proceed to regularize the divergent expression in \eqref{SEpiF1}. Although this formula reduces to the transverse part of the vacuum term at vanishing $eB$, subtracting such a vacuum term is insufficient to eliminate divergences at all orders — a feature that might be related to the non-renormalizability of the Weinberg model. The residual divergence remaining after the subtraction is logarithmic and proportional to $eB$, then it can be fully absorbed via the renormalization of the prefactor that appears in front of the $eB$ term from the free inverse pion propagator. In a word, if we impose the renormalization condition that the inverse pion propagator retains its free form, $-p_0^2+|eB|+m_{\pi^\pm}^2$, in the small magnetic field limit $eB\sim 0$, the self-energy \eqref{SEpiF1} becomes renormalizable and can be properly renormalized without introducing any artificial cutoff as
\bea
\Pi_{\pi^\pm}({p}_\parallel)
&=&-{{g_\rho^2\over8\pi^2}}\int_0^\infty\!\! {\di s\over s\!+\!t}\int_0^\infty\!\! \di t
\left[\left({{1\!-\!e^{-2|eB|s}\over 2|eB|}\!+\!t}\right)^{-2}\!-\!\left(s\!+\!t\right)^{-2}\left(1+{3s+t\over s+t}|eB|s\right)e^{-|eB|s}\right]e^{-s\,M_{\rho^+_\uparrow0}^2-t\,m_{\pi^0}^2+{s\,t\over s+t}{p}_\parallel^2}\nonumber\\
&=&-{{g_\rho^2\over16\pi^2}}\int_0^\infty {\di s\over s^2}\int_{-1}^1 \di u
\left[\left({{1\!-\!e^{-2|eB|s{1+u\over 2}}\over 2|eB|s}\!+\!{1\!-\!u\over 2}}\right)^{-2}\!-\left(1+{(1+u)(2+u)\over2}|eB|s\right)e^{-|eB|{1+u\over 2}s}\right]\nonumber\\
&&\ \ \ \ \ \ \ \ \ \ \ \ \ \ \ \ \ \ \ \ \ \ \ e^{-s\big({1+u\over 2}\,M_{\rho^+_\uparrow0}^2+{1-u\over 2}\,m_{\pi^0}^2-{1-u^2\over 4}{p}_\parallel^2\big)}.\label{SEpiF2}
\eea
Recalling that $M_{\rho^+_\uparrow0}^2\equiv -eB+m_{\rho^+_\uparrow0}^2$, the subtraction term in \eqref{SEpiF2} actually only contains $eB$-linear terms for a constant $m_{\rho^+_\uparrow0}^2$. As the term in the square bracket is of order $o((|eB|s)^2)$ around $|eB|s\sim0$, the integral is indeed convergent and the leading contribution is of order $o(|eB|^2)$ for a small $|eB|$. In the large $eB$ limit, the subtraction term is exponentially suppressed, and the relation ${1-e^{-2|eB|s}\over 2|eB|}\rightarrow {1\over 2|eB|}$ holds. As a result, the self-energy in \eqref{SEpiF2} equals that in \eqref{SEpiF1}, and completing the proper time integrals in the last line of \eqref{SEpiF1} yields exactly the lowest energy given in \eqref{SEpi0}. 

In the three-flavor case, the self-energies of $\pi^\pm$ and $K^\pm$ follow \eqref{SEpi3f} and \eqref{SEK3f}, respectively, as
\bea
\Pi_{\pi^\pm}({p}_\parallel)
&=&-{{g_\rho^2\over16\pi^2}}\int_0^\infty {\di s\over s^2}\int_{-1}^1 \di u
\left[\left({{1\!-\!e^{-2|eB|s{1\!+\!u\over 2}}\over 2|eB|s}\!+\!{1\!-\!u\over 2}}\right)^{-2}\!-\left(1+{(1+u)(2+u)\over2}|eB|s\right)e^{-|eB|{1+u\over 2}s}\right]\nonumber\\
&&\ \ \ \ \ \ \ \ \ \ \ \ \ \ \ \ \ \ \ \ \ \ \ \left[e^{-s\big({1+u\over 2}\,M_{\rho^+_\uparrow0}^2+{1-u\over 2}\,m_{\pi^0}^2-{1-u^2\over 4}{p}_\parallel^2\big)}+{1\over2} e^{-s\big({1+u\over 2}\,M_{K^{*+}_\uparrow0}^2+{1-u\over 2}\,m_{K^0}^2-{1-u^2\over 4}{p}_\parallel^2\big)}\right],\label{SEpi3fF}\\
\Pi_{K^\pm}({p}_\parallel)
&=&-{{g_\rho^2\over32\pi^2}}\int_0^\infty {\di s\over s^2}\int_{-1}^1 \di u
\left[\left({{1\!-\!e^{-2|eB|s{1\!+\!u\over 2}}\over 2|eB|s}\!+\!{1\!-\!u\over 2}}\right)^{-2}\!-\left(1+{(1+u)(2+u)\over2}|eB|s\right)e^{-|eB|{1+u\over 2}s}\right]\nonumber\\
&&\ \ \ \ \ \ \ \ \ \ \ \ \ \ \left[e^{-s\big({1+u\over 2}\,M_{\rho^+_\uparrow0}^2+{1-u\over 2}\,m_{K^0}^2-{1-u^2\over 4}{p}_\parallel^2\big)}+{1\over2}\sum_{S^0=\pi^0}^{\eta,\eta'}\alpha_{S^0} e^{-s\big({1+u\over 2}\,M_{K^{*+}_\uparrow0}^2+{1-u\over 2}\,m_{S^0}^2-{1-u^2\over 4}{p}_\parallel^2\big)}\right].\label{SEK3fF}
\eea
And the lowest energies of $\pi^\pm$ and $K^\pm$ can be solved self-consistently from the equations \eqref{SCEpi} and \eqref{SCEK} by insetting the full self-energies \eqref{SEpi3fF} and \eqref{SEK3fF}.
\end{widetext}

\subsection{Numerical results} \label{FLLN}
\begin{figure}[!htb]
	\centering
	\includegraphics[width=0.42\textwidth]{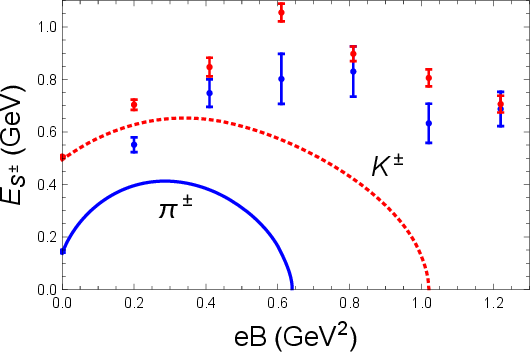}
	\caption{The lowest energy of $\pi^\pm$ and $K^\pm$, $E_{\pi^\pm}$ (black solid) and $E_{K^\pm}$ (red dotted), as functions of $eB$ with the full-Landau-level self-energies Eq.~\eqref{SEpi3fF} and Eq.~\eqref{SEK3fF}, respectively. The blue and red biased lattice data are for  $\pi^\pm$ and $K^\pm$, respectively~\cite{Ding:2026qzu}.}\label{PKF}
\end{figure}

\begin{figure}[!htb]
	\centering
	\includegraphics[width=0.42\textwidth]{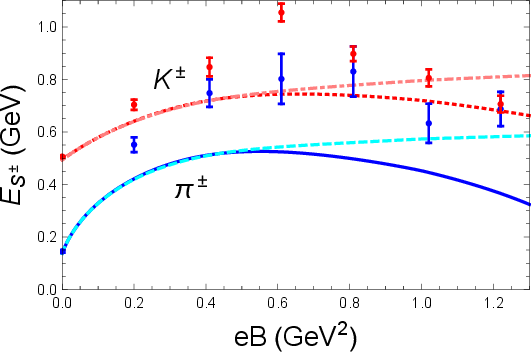}
	\caption{The lowest energy of $\pi^\pm$ and $K^\pm$, $E_{\pi^\pm}$ and $E_{K^\pm}$, as functions of $eB$ with the full-Landau-level self-energies Eq.~\eqref{SEpi3fF} and Eq.~\eqref{SEK3fF}, respectively. Two forms of running couplings are considered for comparison, that is, $g_{\rho1}^2(eB)=6.07^2/(1+0.9\tanh{eB\over \Lambda})$ (blue solid for $\pi^\pm$ and red dotted for $K^\pm$) and $g_{\rho2}^2(eB)=6.07^2/(1+0.9\ln(1+{eB\over \Lambda}))$ (cyan dashed for $\pi^\pm$ and pink dashdotted for $K^\pm$) with $\Lambda=0.3\,{\rm GeV}^2$. The blue and red biased lattice data are for  $\pi^\pm$ and $K^\pm$, respectively~\cite{Ding:2026qzu}.}\label{PKF2}
\end{figure}

In the three-flavor case, the lowest energies of $\pi^\pm$ and $K^\pm$ are demonstrated in Fig.~\ref{PKF} for the full-Landau-level calculations, in comparison with the lattice results~\cite{Ding:2026qzu}. As we can see, compared to Fig.~\ref{PKLE}, the main features are the same thus confirm the qualitative analysis with the lowest energies, but the peaks and unstable points are shifted to much larger $eB$.
In Fig.~\ref{PKF2}, we show the full-Landau-level calculations with two forms of running couplings, $g_{\rho1}^2(eB)=6.07^2/(1+0.9\tanh{eB\over \Lambda})$ and $g_{\rho2}^2(eB)=6.07^2/(1+0.9\ln(1+{eB\over \Lambda}))$ with $\Lambda=0.3\,{\rm GeV}^2$. As we can see, in the former case, the peaks are shifted to $|eB|\sim 0.6\,{\rm GeV}^2$ and the unstable points are greatly postponed; while in the latter case, there are no peaks at all in the considered region and the lowest energies increase very slowly at larger $eB$. Recalling that the auxiliary function $\xi(|eB|)\approx {g_\rho^2\over4\pi^2}\ln (|2eB|/{M}_{\rho^+_\uparrow0}^2)$ at a large enough $eB$, it would approach a constant, $1.036$, in the latter case for a slowly varying ${M}_{\rho^+_\uparrow0}^2$. Then, it turns out that the lowest energy, $M^2_{\pi^\pm_0}=-0.036|eB|+m_{\pi^\pm}^2$, weakly depends on $eB$ -- this can roughly explain why $E_{\pi^\pm}$ and $E_{K^\pm}$ vary very slowly in the larger $eB$ region. Since $M^2_{\pi^\pm_0}$ is expected to decrease with $eB$ in the lowest-Landau-level approximation, the increasing feature just indicates that the upper limit of magnetic field, $\sim 1\,{\rm GeV}^2$, is not large enough to ensure the approximation. But we are not able to extend the calculations to much larger $|eB|$ to check that, since the lattice data only cover the magnetic field region $|eB|\leq 1.22\,{\rm GeV}^2$. At least, well within the lattice region, the running coupling constant $g_{\rho}^2(eB)$ should decrease more slowly than $1/\ln{eB} $ at larger $eB$ in order to produce peak structures. This is somehow like fine-tuned.

\section{Summary and discussions}\label{summary}

In this work, we try to solve the quite long-standing puzzle raised by lattice QCD simulations: The lowest energies of charged pseudoscalar mesons, $\pi^\pm$ and $K^\pm$, decrease with stronger magnetic field. To do that, we choose to work within the two- and three-flavor Weinberg models, where intrinsically pseudoscalar and vector mesons couple with each other.
\begin{itemize} [leftmargin=1em]
\item Firstly, at the lowest-Landau-level approximation, it is consistently found that the lowest energies of $\pi^\pm$ and $K^\pm$ indeed decrease with stronger magnetic field, a behavior driven by the self-energy corrections introduced through neutral pseudoscalar-charged vector loops. However, the peaks are far smaller than the lattice predictions; even more critically, the lowest energies would decrease to zero at stronger magnetic field, a signature of instabilities.
\item Secondly, we explore the lowest energies of $\pi^\pm$ and $K^\pm$ by adopting the full-Landau-level propagators for charged vectors: While the features of peaks and instabilities remain in the three-flavor case, the corresponding magnetic fields both shift to higher magnitudes‌. Consequently, the downward trend of energy is unavoidable at stronger magnetic fields when the coupling constant $g_\rho$ is held fixed, which aligns with the intuitive expectation that the lowest-Landau-level approximation becomes progressively more valid as $eB$ increases.
\item Thirdly, two forms of running coupling constant are considered to examine how the features of the lowest energies are modified. It turns out that even for a running coupling constant that decreases as slowly as $1/\ln eB$ at large $eB$, no peak structures can be reproduced within the magnetic field range covered by lattice QCD. For a running coupling constant decreasing as slowly as $1/\tanh eB$ at larger $eB$, the peak structures are well reproduced and no instabilities are observed in the considered magnetic field region.
 \end{itemize}
Recalling that the lowest energies of charged pseudoscalars are enhanced by quark-antiquark loop corrections~\cite{Cao:2015xja,Ke:2026npb}, our results provide explicit support for the conjecture proposed in Ref.~\cite{Kojo:2026soi}: A charged pseudoscalar gradually evolves from a two-quark state to a tetraquark state composed of a neutral pseudoscalar and a charged vector meson as the magnetic field increases. In this context, the peaks in the lowest-energy spectra of charged pseudoscalars can be identified as the transition points.\\

Though the interactions with neutral pseudoscalars and charged vectors provide valuable insights into the decreasing behavior of the lowest energies $\pi^\pm$ and $K^\pm$ at larger $eB$, a set of problems still remain:
\begin{itemize} [leftmargin=1em]
\item[1.] Why does the peak structure of $K^\pm$ exhibit a much steeper profile compared with that of $\pi^\pm$ in lattice QCD simulations ? The former peak even significantly exceeds the free-particle result. This might indicate that the enhancement originating from quark-antiquark loops plays a critical role for charged pseudoscalars across the peak region, which is not incorporated in the simple mesonic Weinberg model.
\item[2.] Even with a properly tuned running coupling constant, it remains impossible to simultaneously reproduce the observed peak structures and eliminate the unphysical instabilities within the Weinberg model framework. Since $\pi^\pm$ or $K^\pm$ superfluidity is not physically expected to emerge in a pure magnetic field, this inconsistency constitutes a severe theoretical issue. Again, the intrinsic quark-antiquark substructure may render the dynamics of neutral pseudoscalars not fully three-dimensional in the strong-field regime. 
\end{itemize}
 Therefore, in the future, the quark-antiquark polarization effect should be explicitly incorporated into the meson sector to achieve a better description of lattice data within the Weinberg model. In parallel, QCD-based nonperturbative functional approaches~\cite{Fu:2019hdw,Gao:2020fbl} can be employed to systematically verify whether the neutral pseudoscalar-charged vector loop contributions are sufficient to fully account for the decreasing features of the lowest energies $\pi^\pm$ and $K^\pm$.

\section*{Acknowledgment}
 G.C. thanks Toru Kojo and Hengtong Ding for helpful discussions. G.C. is funded by the National Natural Science Foundation of China with Grant Nos. 12447102 and 12575152, and the Natural Science Foundation of Guangdong Province with Grant No. 2024A1515011225.

\end{document}